%% file: main.tex
\documentclass[letterpaper,11pt]{article}
\usepackage{graphicx} 
\usepackage{rotating} 
\usepackage{lscape} 
\usepackage{xcolor}
\usepackage{amssymb}
\usepackage{enumitem}
\usepackage{array}
\usepackage{comment}

\title{No More Trade-Offs.\\GPT and Fully Informative Privacy Policies}

\usepackage{authblk}
\author{Przemysław Pałka$^{1,}$\thanks{The work of P.~Pałka, between September 3 and October 3, 2023, was funded by and conducted within the frame of the Visiting Researcher program at the Department of Legal Studies,  University of Bologna, Italy. The research prior to and following this period was funded by the National Science Center, Poland, grant no. 2022/45/B/HS5/01419.}, Marco Lippi$^{2,}$\thanks{Research partly funded by PNRR - M4C2 - Investimento 1.3, Partenariato Esteso PE00000013 - ``FAIR - Future Artificial Intelligence Research'' - Spoke 8 ``Pervasive AI'', funded by the European Commission under the NextGeneration EU programme.},\\\vspace{-5pt}Francesca Lagioia$^{3,4,\dagger,}$\thanks{European Research Council (ERC) Project ``CompuLaw'' (Grant Agreement No 833647) under the European Union's Horizon 2020 research and innovation programme.}, Rūta Liepiņa$^{3,\ddagger}$, Giovanni Sartor$^{3,4,\dagger,\ddagger}$}
\date{
$^1$ Jagiellonian University, przemyslaw1.palka@uj.edu.pl\\
$^2$ University of Modena and Reggio Emilia\\
$^3$ Alma Mater Studiorum University of Bologna\\
$^4$ European University Institute\\
}

\begin{document}

\maketitle

\begin{abstract}
The paper reports the results of an experiment aimed at testing to what extent ChatGPT 3.5 and 4 is able to answer questions regarding privacy policies designed in the new format that we propose. In a world of human-only interpreters, there was a trade-off between comprehensiveness and comprehensibility of privacy policies, leading to the actual policies not containing enough information for users to learn anything meaningful. Having shown that GPT performs relatively well with the new format, we provide experimental evidence supporting our policy suggestion, namely that the law should require fully comprehensive privacy policies, even if this means they become less concise. 
\end{abstract}

\newpage
\section{Introduction}

Imagine you just listened to a song or ordered food using an app. Imagine asking your smartphone: ``Hey Skynet, could you tell me the names of all corporations who will know that I just listened to \textit{Italodisco}, or ordered tofu,  and list all the purposes for which they will use this data?''\footnote{These are no (longer) trivial matters; in a ML-powered inference economy~\cite{solow2022information} such data can be proxy for estimating one's race~\cite{marshall2018s} or political convictions~\cite{dellaposta2015liberals}.} As a consumer, you have the right to know this~\cite{calo2011against,kaminski2019right,mahieu2020harnessing}. Such knowledge is a precondition for choosing whether and how to use various services~\cite{reidenberg2015disagreeable}. 
And yet, in 2023, asking such a question and receiving a straightforward answer is not possible.

The reason behind this impossibility is not (only) technological. Due to the spread of large language models (LLMs), machines are nowadays able to read and comprehend natural language at a level that was unthinkable only a few years ago~\cite{radford2019language}. Therefore, we posit that the reason is chiefly that the information necessary to answer such questions \textit{is not} publicly available in privacy policies~\cite{palka2023}. And that the law is partly to blame for that.

The law\footnote{In this paper we focus on EU law, i.e. the GDPR, and account for the American law.} requires privacy policies to be simultaneously \textit{comprehensive}, i.e., to contain information about all the data categories processed, purposes of use, and their recipients; and \textit{comprehensible}, i.e., to be concise, written in simple language, and easy to understand. In the world of human-only interpreters, there is a trade-off between the two: a shorter and simpler policy would be less comprehensive, and \textit{vice versa.} The comprehensibility requirement has been advocated for in scholarship, calling for ever more simplification~\cite{calo2011against,emami2021informative,kelley2009nutrition,waldman2018privacy}. 

In this paper, we challenge the conventional wisdom and take the opposite approach. We argue that the law should focus on requiring the privacy policies to be fully comprehensive, \textit{even if} the cost of comprehensiveness is lowering  comprehensibility of such documents for individual consumers. The reason is simple: soon anyone could engage in simplification of, or information retrieval from, long privacy policies, 
as such efforts can be increasingly automated. 
However, for these summaries or answers to be meaningful, they need to be based on comprehensive information about corporate practices concerning consumers' personal data. 

This policy suggestion assumes the capacity of AI to process fully comprehensive privacy policies.
Hence, we decided to verify that premise. In this paper, we report some preliminary results of our first experiment, aimed at testing the extent to which tools like ChatGPT are already capable of answering consumer questions based on fully comprehensive privacy policies (and \textit{not} the \textit{existing} privacy policies, as tested in~\cite{harkous2018polisis,ravichander2019question,tang2023policygpt}). To do so, we have written a mock privacy policy of ``Orderoo Inc.,'' a company similar to DoorDash, UberEats, etc., in the format we propose as fully comprehensive. The format is presented below in Appendix~\ref{sec:appendixC}, and the full privacy policy, with all the experiment documentation, made available on GitHub.\footnote{~https://github.com/ruutaliepina/full-privacy}

We design our experiment as a proof-of-concept across four different settings, i.e., using short and long prompts, for ChatGPT 3.5 and 4. The results are interesting and promising, as they clearly show how LLMs are able to digest the information contained in long documents, and to answer complex questions that need to connect several paragraphs. Nevertheless, we note how the provided answers sometimes contain mistakes, which poses an additional challenge from the perspective of computer science, to what extent the output of these models can be further controlled and post-processed in order to make reliable information available to the end user.


The paper proceeds in five steps. First, we discuss the related work, and second, the legal background.
Third, we describe what a fully comprehensive privacy policy could look like. Fourth, we report the results of our preliminary experiments. 
Finally, we discuss implications for further research in computer science, law, and policy. 

\section{Related Work}

Empirical scholars have shown that consumers do not read privacy policies~\cite{bakos2014does, obar2020biggest} and, when they do, they often misunderstand them~\cite{reidenberg2015disagreeable}. One study estimated that reading the privacy policies of all the websites one uses would take hundreds of hours each year~\cite{mcdonald2008cost}. As a solution to this problem, numerous researchers have suggested making privacy policies even shorter or easier to understand, e.g., by using the so-called ``privacy nutrition labels''~\cite{calo2011against,emami2021informative,kelley2009nutrition,waldman2018privacy}. However, it has been also 
shown that even simplified privacy policies have a close-to-zero effect on consumers' understanding of corporate data practices or their rights~\cite{ben2016simplification}.

This might suggest that the problem with privacy policies is not (only) their form but also their content. Put simply: privacy policies often \textit{do not contain} the information necessary for consumers -- even the hypothetical ones with time and expertise to read them -- to learn how exactly their data will be processed~\cite{contissa2018claudette}. A recent study~\cite{palka2023} highlighted how, by using vague terms for data categories (``we collect data about your \textit{use of our service}") or purposes of processing (``we use your data to \textit{improve our service}"),  many privacy policies leave corporations the freedom to engage in data practices that consumers cannot reasonably expect. The same study demonstrated that many privacy policies do not explain what data will be used for what purpose or shared with what recipient~\cite{palka2023}.

 From a technical perspective, several studies have shown that natural language processing (NLP) tools can be used to summarize privacy policies~\cite{zaeem2018privacycheck}, answer users' questions~\cite{ravichander2019question}, or represent the contents of the policies in a user-friendly form~\cite{harkous2018polisis}. However, the usefulness and the practical value of such automated representation, summarizing, and question-answering will be only as good and specific as the information \textit{actually included} in the privacy policies. As of today, the value of this information is limited.

The recent breakthroughs in LLMs have created new opportunities for analyzing texts through simple prompt-based queries. The LLMs use transformer architecture and have been trained on vast amounts of general data such as webpages, books, articles and cover a wide range of topics~\cite{vaswani2017attention}. Legal language is highly specialised and normally general language models perform poorly without additional domain specific fine-tuning~\cite{chalkidis2021lexglue}. While these developments are fairly recent, some studies testing the abilities of LLMs in legal analytics have already emerged~\cite{savelka2023can}. It has been shown that GPT models (3, 3.5, and 4) can answer the US~\cite{katz2023gpt} and Japanese~\cite{yu2023exploring} bar exam questions on par with the examinees. These tests include various reasoning tasks such as rule application, legal opinion drafting, and multiple choice questions. As for privacy policy analysis, Tang et al.~\cite{tang2023policygpt} experimented with ChatGPT and GPT4 to improve information  extraction from privacy policy relevant to the GDPR -- e.g., ``how and why a service provider collects user information,'' ``how long user information is stored.'' The reported results of their PolicyGPT show a significant increase in retrieval metrics as compared to the older models, which used expert annotated datasets. 

\section{Legal Requirements for Privacy Policies: Comprehensiveness vs. Comprehensibility}

The law assigns the drafters of privacy policies a difficult task. On the one hand, it requires the privacy policies to be \textit{comprehensive}, i.e., contain all the information relevant to a consumer pondering whether to use a service. On the other, it demands  privacy policies be \textit{comprehensible}, i.e., written in simple language and easy to understand for consumers.

In the United States, online consumer privacy is governed by the so-called ``notice and choice" model~\cite{richards2018pathologies}. Within its logic, companies who wish to collect and use personal data should make privacy policies available to consumers (``notice'') who, in turn, can decide whether such data practices are acceptable to them (``choice''). Whether companies actually live up to their promises is policed by both contract law~\cite{bar2017searching} and public enforcement by the Federal Trade Commission~\cite{solove2014ftc} and state Attorneys General~\cite{citron2016privacy}. The latter can also issue guidelines. For example, in 2014, then-AG of California, Kamala Harris~\cite{harris2014ag}, stated that privacy policies ``should provide a comprehensive overview of [...] practices regarding the collection, use, sharing and protection of personally identifiable information" and recommended that drafters of privacy policies ``use plain, straightforward language. Avoid technical or legal jargon. Use short sentences."\footnote{~\cite{harris2014ag}, pp 9-10.}


In the European Union, the omnibus General Data Protection Regulation\footnote{Regulation (EU) 2016/679, OJ L 119, 4.5.2016, p. 1–88, hereinafter ``the GDPR."} governs online consumer privacy, among other areas of life~\cite{yeung2022demystifying}. It requires that data controllers provide consumers with detailed information about their data practices,\footnote{The GDPR, arts. 13-14.} yet do so in ``in a concise, transparent, intelligible and easily accessible form, using clear and plain language."\footnote{The GDPR, art. 12.1.} The guidelines endorsed by the European Data Protection Board\footnote{Art. 29 WP, Guidelines on transparency under Regulation 2016/679, last Revised and Adopted on 11 April 2018, available at: https://edpb.europa.eu/our-work-tools/our-documents/guidelines/transparency. Hereinafter ``The Transparency Guidelines."} -- a body comprised of the Supervisory Authorities from all the EU Member States -- provide some examples of how to square the two: 
\\

\noindent\textbf{Poor Practice Examples}
\vspace{0pt}
\begingroup
\addtolength\leftmargini{-0.25in}
\begin{quotation}
\noindent``We may use your personal data to develop new services'' (as it is unclear what the ``services'' are or how the data will help develop them);

\noindent``We may use your personal data for research purposes'' (as it is unclear what kind of ```research'' this refers to); and

\noindent``We may use your personal data to offer personalised services'' (as it is unclear what the ``personalisation'' entails).
\end{quotation}
\endgroup

\vspace{0pt}
\noindent\textbf{Good Practice Examples}
\vspace{0pt}
\begingroup
\addtolength\leftmargini{-0.25in}
\begin{quotation}
\noindent ``We will retain your shopping history and use details of the products you have previously purchased to make suggestions to you for other products which we believe you will also be interested in'' (it is clear what types of data will be processed, that the data subject will be subject to targeted advertisements for products, that their data will be used to enable this).\footnote{The Transparency Guidelines, p. 9.}
\end{quotation}
\endgroup

This looks clear and simple when presented as an isolated example. However, when one considers the sheer volume of data processed by online companies nowadays, the multitude of purposes for which different categories of data are used, or the range of entities with whom data is shared, the \textit{comprehensive} privacy policy quickly becomes inflated. 
In the world of human-only readers, there is an inherent trade-off between comprehensiveness and comprehensibility. Admittedly, corporations have business interests not to disclose too much, like trade secrets or good PR. However, by requiring that privacy policies were simultaneously comprehensive \textit{and} comprehensible, the law has given the corporations \textit{an excuse.} They can always defend themselves by stating ``but we simply tried to make the privacy policy shorter!'' Fortunately, human readers could soon be significantly assisted by AI. 


\section{The Legal Proposal: Fully Comprehensive Privacy Policies}

Our policy suggestion aims to create an informational environment where AI's full potential in assisting consumers can be realized. We propose that the law should require corporations to disclose \textit{fully comprehensive} privacy policies, ideally in a standardized form, and relax the \textit{comprehensibility} requirements. Such notices can be later processed by other parties to suit particular consumers' informational needs, potentially via automated means. 

Under the GDPR's transparency requirements (which we consider a good start) the \textit{unit  of information} that must be disclosed is \textit{an act of processing} of personal data. An act of processing means any action that a data controller takes regarding personal data.\footnote{GDPR art. 4.1.}
Thus, transparency at the stage in which personal data are collected requires the consumers to be informed, among others,  about the following: 
\begin{enumerate}
    \item \textit{Categories of personal data concerned} (e.g., an email address);
     \vspace{-0.75em} 
    \item  \textit{Purpose of processing for which each category of data are intended} (e.g., issuing receipts)
   \vspace{-0.75em} 
    \item \textit{Legal basis} for each data  processing\footnote{Art. 6.1. GDPR lists six possible bases for processing: (a) consent, (b) contractual necessity, (c) legal obligation, (d) subject's vital interest, (e) public task, or (f) controller's legitimate interest. In the analyzed context, (a), (b), (c) and (f) are relevant. Whenever a data controller relies on (f), they should specify what legitimate interest they have in mind, to enable the proportionality assessment(Article 6(1)(f)).} (e.g., contractual necessity)
  \vspace{-0.75em} 
    \item \textit{Storage period} or, when it is not possible, the \textit{criteria to determine such a period} (e.g., till account closure), and with regard to each data and purpose 
    \vspace{-0.75em} 
    \item \textit{Recipients} or at least the \textit{categories of recipients} (
    e.g., a cloud provider), as well as their
    \vspace{-0.75em} 
    \begin{enumerate}
        \item Role (controller or processor)
        \vspace{-0.25em} 
        \item Purpose of sharing (e.g., data backup)
        \vspace{-0.25em} 
        \item Legal basis of sharing (e.g., legitimate interest)
    \end{enumerate}
\end{enumerate}

 In our empirical work on privacy policies~\cite{contissa2018claudette,contissa2018automated,liepin2019gdpr,palka2023} we have not yet encountered a single policy that is compliant with all these requirements.
 This may be due to different reasons, one of them probably being that 
 such a document would not meet the concise  and comprehensibility requirements. Note how a single category of data (e.g., an email address) might be used for several purposes, e.g., sending receipts, marketing information, serve as an identifier, etc. Each of these purposes must be described as a separate unit of information, as they might have different legal bases. 

 Such information can be represented in several different ways. On the one had, there are various machine-readable formats~\cite{kumaraguru2007survey,tondel2012towards} (though scholarly attempts at constructing those seem to have culminated about a decade ago), potentially based on ontologies~\cite{gharib2020ontology,palmirani2018pronto}. On the other hand, there are human-readable formats, i.e., texts or tables. However, the advances in NLP~\cite{zhong2020does} contribute to the blurring of the machine-/human-readable distinction. 

 In this paper, we propose two formats for \textit{fully comprehensive} privacy policies. First, we suggest creating a spreadsheet with 8 columns: (A) category identifier, (B) data type, (C) source, (D) purpose, (E) purpose explanation, (F) legal basis, (G) legal basis explanation, and (H) storage period. The number of rows should equal the number of purposes of processing for each data type, as these might have different legal bases and be stored for different periods. In addition, a second sheet describes the sharing of data with other parties, in a similar form. We illustrate how this could look like in Appendix \ref{sec:appendixB}. The value of this approach is in its clarity and logical structure of all the data points without the added noise.

 Second, exactly the same information can be represented as solid text. Here, each paragraph corresponds to one data type and lists the source and all the purposes with their legal bases. Then, it lists all the recipients of personal data and identifies their role (controller or processor), the purpose of sharing, and its legal basis. We illustrate how this could look like in Appendix \ref{sec:appendixC}. We conduct the experiment on the solid text version.

\section{The Experimental Methodology}

We run an experiment as a proof-of-concept of our methodology. Our goal is to test to what extent the consumer-facing versions of ChatGPT 3.5 and 4 may correctly answer user questions regarding the contents of fully comprehensive privacy policies. To do so, we created a mock privacy policy of Orderoo Inc., a food delivery company similar to DoorDash and Uber Eats.\footnote{Note that the mock policy is in the fully comprehensive \textit{format} but could be longer if we accounted for more categories of data.}


We identified the following six questions to be used for the test:
\begin{enumerate}
    \item [Q1:] What data does Orderoo process about me?
     \vspace{-0.75em} 
    \item [Q2:] For what purposes does Orderoo use my email address?
   \vspace{-0.75em} 
    \item [Q3:] Who does Orderoo share my geolocation with?
  \vspace{-0.75em} 
    \item [Q4:] What types of data are processed on the basis of consent, and for what purposes? 
    \vspace{-0.75em} 
    \item [Q5:] What data does Orderoo share with Facebook?
    \vspace{-0.75em} 
   \item [Q6:] Does Orderoo share my data with insurers?
\end{enumerate}

The three initial questions are rather straightforward and aimed to test the model's ability to retrieve the relevant information and connect different informative elements. The answer to Q1 can be found at the beginning of each paragraph of the policy, while the answers to Q2 and Q3 are contained in a single paragraph. The last three questions are more complicated. The answers to Q4 and Q5 are spread throughout the entire document. Q6 requires a negative answer, i.e., the policy does not mention any data sharing with insurers. The latter is mostly aimed to test potential hallucinations.  

As noted above, for the purpose of this work, we tested ChatGPT versions 3.5 and 4, on Sep 30-Oct 2, 2023, while located in Bologna, Italy. We defined two prompts, one short and one long, the latter aimed at mitigating hallucinations (i.e., making answers up)~\cite{azamfirei2023large,bang2023multitask}, unwarranted simplification (i.e., listing only some answers, preceded by phrases like ``for example'' or ``such as''), and unwarranted text generation (i.e., adding information that the user did not ask for, e.g., advice that one should carefully read the laws, etc.).
%
%
%
As an example consider the following:

\begingroup
\addtolength\leftmargini{-0.1in}
\begin{quotation}
\noindent \textbf{Short prompt:} \textit{What data does Orderoo process about me?}
\end{quotation}
\endgroup

\begingroup
\addtolength\leftmargini{-0.1in}
\begin{quotation}
\noindent \textbf{Long prompt:} \textit{What data does Orderoo process about me? In answering the question please rely solely on the information included in the text and not your knowledge from other sources; please read the document carefully and mention everything, do not omit any information included in the text; please do not shorten or simplify the answers by inserting elements like ``for example'', ``including'' or ``such as'', please limit your answer strictly to what I am asking about and refrain from giving me advice or informing me about things I have not asked about.}
\end{quotation}
\endgroup

Hence, we considered four settings: (1) GPT 3.5, short prompt; (2) GPT 3.5, long prompt; (3) GPT 4 short prompt; (4) GPT 4 long prompt.
For each setting, we repeated the test five times with two independent accounts, for a total of ten runs.
Each chat began with the same statement, i.e., \textit{Hi, I will copy-paste a document here and then ask you some questions about its contents, is that ok?}, followed by the entire mock privacy policy of Orderoo Inc. 
For the purpose of evaluation, we consider the following four cases: (i) correct answer; (ii) hallucination; (iii) false positive (when the answer mentioned information that clearly was taken from the document, though wrong); and (iv) false negative (when the answer failed to mention something). We also decided, whenever the answer was not correct, to prompt GPT again with a generic ``redo it'' question (i.e. \textit{Are you sure? Please try again}), thus recording whether this led to improved performance or not.
All the answers were manually checked, and the whole set of conversations is available at the aforementioned repository.


Our initial hypotheses were the following:
\begin{enumerate}
    \item[H1:] There will be a problem with hallucination, especially when asking about Facebook (as GPT ``knows'' a lot about it from other sources)
   \vspace{-0.75em} 
    \item[H2:] There should be no problem with false positives, and false negatives should occur sporadically, due to unwarranted simplification
  \vspace{-0.75em} 
    \item[H3:] GPT-4 will do better than GPT-3.5 on all the questions
    \vspace{-0.75em} 
    \item[H4:] Longer prompts will lead to better answers than shorter ones
\end{enumerate} 

\section{Experimental Results and Discussion}

\begin{table}
\begin{center}
\begin{tabular}{lcccccc}
  & Q1 & Q2 & Q3 & Q4 & Q5 & Q6 \\
  \hline
  GPT-3.5 (S) & 10 & 10 & 10 & 0 & 0 & 10 \\ 
  GPT-3.5 (L) & 10 & 10 & 10 & 0 & 0 & 10 \\ 
  GPT-4 (S) & 10 & 10 & 7 & 4 & 10 & 10 \\ 
  GPT-4 (L) & 10 & 10 & 7 & 2 & 10 & 10 \\ 
\end{tabular}
\end{center}
\caption{Results obtained by GPT-3.5 and GPT-4, with short (S) and long (L) prompts, respectively. We report the number of correct answers out of 10 runs, obtained with two different accounts (5 runs each). The complete list of results is reported in Appendix~\ref{sec:appendixA}.\label{tab:results_summary}}
\end{table}

Table~\ref{tab:results_summary} reports the results obtained across the different settings. 
Notably, in none of the settings did GPT manage to answer all six questions correctly. However, several interesting observations can be made based on these results. Significantly, results were not always consistent across the runs. For Q3 and Q4, in fact, one of the two accounts obtained different answers when questioning GPT-4 both with short and long prompts. 

Regarding H1, none of the experiments reported hallucination. This suggests that, when asked questions about a document uploaded by the user, the hallucination problem is less pertinent than with open-ended questions that rely on GPT's ``own knowledge.'' Hence, H1 has been falsified.

Concerning H2 and H3, the only instances of false positives occurred in Q3, with GPT-4 (though only for one account). The system listed one additional entity with whom Orderoo allegedly shares geolocation, namely ``Cloud711''. This is the name of a company made up by us, mentioned repeatedly in the document, so this was not a hallucination. Moreover, every data type other than geolocalization is shared by Orderoo with Cloud711, according to the policy. Thus, in this case, the system just wrongly interpreted the text. All other runs got this correct. False negatives, instead, consistently appeared in Q3 and Q4 for GPT-3.5 (all 10 runs, both short and long prompts) and also, though less frequently, for GPT-4 (6 and 8 cases for short and long prompts in Q4, respectively; only once in Q3).
Therefore, given all this evidence, also H2 and H3 have been partly falsified. 

Remarkably, all runs correctly answered Q1 and Q2 
and Q6 (where a negative answer was correct). This suggests that having even longer privacy policies, where additional sections are organized not only around data categories but also around purposes and legal bases, could help. However, this would mean that the documents will be significantly longer (more processing power needed) and come with the additional need to ensure consistency throughout the document. 
Surprisingly, the longer prompts had no effect on GPT 3.5. For GPT 4, instead, the results were interesting: the additional instruction led to \textit{more} false negatives. In this sense, H4 has been falsified. 

The good performance of GPT-4 when answering questions about the fully comprehensive privacy policy (if we consider the majority of cases, 5 correct answers out of 6 questions) suggests that our policy proposal should be seriously considered by lawmakers or even privacy-sensitive corporations acting on their own motion. This proposal challenges the conventional wisdom that privacy policies should be made shorter rather than longer and simpler rather than more complex~\cite{calo2011against,emami2021informative,kelley2009nutrition,waldman2018privacy}. However, for this reform to indeed meaningfully affect the position of consumers, further research in computer science is needed.

Some limitations with the use of pretrained systems should be noted. Currently, ChatGPT free version limits the input text to approximately 3000 words. When arguing for more comprehensive privacy policies, the word limit could create problems. For example, during the experiments, we also tested the performance of the system by entering the privacy policy splitting it over two messages. The results were significantly worse compared to the single input -- the answers were based on general information about privacy policies, not the specific mock policy we provided. Moreover, GPT architecture remains opaque and users are given limited control over the results they receive.

\section{Conclusion and Future Work}

In this paper, we suggested that making privacy policies \textit{fully comprehensive} is a pro-consumer move, provided that automated means for question answering are available. We created a mock privacy policy, demonstrating what format we have in mind, and tested the performance of Chat GPT 3.5 and 4, using both short and long prompts. The novelty of this experiment lays in the creation and use of a synthetic privacy policy to test the abilities of the new LLMs. Even though the results where not perfect, the good performance of GPT 4 suggests that this is a promising path. 

As part of experimental work, we plan to expand the mock privacy policy and the number of documents to be assessed,  test different LLMs, including open source LLMs; explore the use of API to create further controls for the quality of the answers, as well as, link the answers to the relevant portions of the text.




\bibliographystyle{plain}
\bibliography{biblio}
\newpage

\appendix

\section*{APPENDIX}
\section{Additional details on experimental results}\label{sec:appendixA}

We hereby report the complete list of the results obtained with the 10 different runs (5 for each of two accounts) across the four considered settings.

\begin{table}[!h]
\begin{center}
\begin{tabular}{c|ccccc|ccccc}
& \multicolumn{5}{c|}{Account \#1} & \multicolumn{5}{c}{Account \#2} \\
& $r_1$ & $r_2$ & $r_3$ & $r_4$ & $r_5$ & $r_1$ & $r_2$ & $r_3$ & $r_4$ & $r_5$ \\
\hline
Q1 & \checkmark & \checkmark & \checkmark & \checkmark & \checkmark & \checkmark & \checkmark & \checkmark & \checkmark & \checkmark \\
Q2 & \checkmark & \checkmark & \checkmark & \checkmark & \checkmark & \checkmark & \checkmark & \checkmark & \checkmark & \checkmark \\
Q3 & \checkmark & \checkmark & \checkmark & \checkmark & \checkmark & \checkmark & \checkmark & \checkmark & \checkmark & \checkmark \\
Q4 & FN & FN & FN & FN & FN & FN & FN & FN & FN & FN \\
Q5 & FN & FN & FN & FN & FN & FN & FN & FN & FN & FN \\
Q6 & \checkmark & \checkmark & \checkmark & \checkmark & \checkmark & \checkmark & \checkmark & \checkmark & \checkmark & \checkmark \\
\end{tabular}
\caption{Results obtained in \textbf{Setting 1}, i.e., with GPT-3.5 with short prompts on the 10 overall runs (5 for two accounts each). A checkmark indicates a correct answer. FN indicates that the answer contains a false negative.\label{tab:results_gpt3.5_short_complete}}
\end{center}
\end{table}

\begin{table}[!h]
\begin{center}
\begin{tabular}{c|ccccc|ccccc}
& \multicolumn{5}{c|}{Account \#1} & \multicolumn{5}{c}{Account \#2} \\
& $r_1$ & $r_2$ & $r_3$ & $r_4$ & $r_5$ & $r_1$ & $r_2$ & $r_3$ & $r_4$ & $r_5$ \\
\hline
Q1 & \checkmark & \checkmark & \checkmark & \checkmark & \checkmark & \checkmark & \checkmark & \checkmark & \checkmark & \checkmark \\
Q2 & \checkmark & \checkmark & \checkmark & \checkmark & \checkmark & \checkmark & \checkmark & \checkmark & \checkmark & \checkmark \\
Q3 & \checkmark & \checkmark & \checkmark & \checkmark & \checkmark & \checkmark & \checkmark & \checkmark & \checkmark & \checkmark \\
Q4 & FN & FN & FN & FN & FN & FN & FN & FN & FN & FN \\
Q5 & FN & FN & FN & FN & FN & FN & FN & FN & FN & FN \\
Q6 & \checkmark & \checkmark & \checkmark & \checkmark & \checkmark & \checkmark & \checkmark & \checkmark & \checkmark & \checkmark \\
\end{tabular}
\caption{Results obtained in \textbf{Setting 2}, i.e., with GPT-3.5 with long prompts on the 10 overall runs (5 for two accounts each). A checkmark indicates a correct answer. FN indicates that the answer contains a false negative.\label{tab:results_gpt3.5_long_complete}}
\end{center}
\end{table}

\begin{table}[!h]
\begin{center}
\begin{tabular}{c|ccccc|ccccc}
& \multicolumn{5}{c|}{Account \#1} & \multicolumn{5}{c}{Account \#2} \\
& $r_1$ & $r_2$ & $r_3$ & $r_4$ & $r_5$ & $r_1$ & $r_2$ & $r_3$ & $r_4$ & $r_5$ \\
\hline
Q1 & \checkmark & \checkmark & \checkmark & \checkmark & \checkmark & \checkmark & \checkmark & \checkmark & \checkmark & \checkmark \\
Q2 & \checkmark & \checkmark & \checkmark & \checkmark & \checkmark & \checkmark & \checkmark & \checkmark & \checkmark & \checkmark \\
Q3 & \checkmark & \checkmark & \checkmark & \checkmark & \checkmark & FP & \checkmark & \checkmark$^*$ & FP$^*$ & \checkmark \\
Q4 & FN & FN & FN & FN & FN & \checkmark & \checkmark & FN & \checkmark & \checkmark \\
Q5 & \checkmark & \checkmark & \checkmark & \checkmark & \checkmark & \checkmark & \checkmark & \checkmark & \checkmark & \checkmark \\
Q6 & \checkmark & \checkmark & \checkmark & \checkmark & \checkmark & \checkmark & \checkmark & \checkmark & \checkmark & \checkmark \\
\end{tabular}
\caption{Results obtained in \textbf{Setting 3}, i.e., with GPT-4 with short prompts on the 10 overall runs (5 for two accounts each). A checkmark indicates a correct answer. FN and FP indicate that the answer contains a false negative or false positive, respectively. $^*$ indicates that the answer changed after the system was asked to rethink the answer. \label{tab:results_gpt4_short_complete}}
\end{center}
\end{table}

\begin{table}[!h]
\begin{center}
\begin{tabular}{c|ccccc|ccccc}
& \multicolumn{5}{c|}{Account \#1} & \multicolumn{5}{c}{Account \#2} \\
& $r_1$ & $r_2$ & $r_3$ & $r_4$ & $r_5$ & $r_1$ & $r_2$ & $r_3$ & $r_4$ & $r_5$ \\
\hline
Q1 & \checkmark & \checkmark & \checkmark & \checkmark & \checkmark & \checkmark & \checkmark & \checkmark & \checkmark & \checkmark \\
Q2 & \checkmark & \checkmark & \checkmark & \checkmark & \checkmark & \checkmark & \checkmark & \checkmark & \checkmark & \checkmark \\
Q3 & \checkmark & \checkmark & \checkmark & \checkmark & \checkmark & \checkmark & FP & FP & FP & \checkmark \\
Q4 & FN & FN & FN & FN & FN & FN & FN & FN & \checkmark & \checkmark \\
Q5 & \checkmark & \checkmark & \checkmark & \checkmark & \checkmark & \checkmark & \checkmark & \checkmark & \checkmark & \checkmark \\
Q6 & \checkmark & \checkmark & \checkmark & \checkmark & \checkmark & \checkmark & \checkmark & \checkmark & \checkmark & \checkmark \\
\end{tabular}
\caption{Results obtained in \textbf{Setting 4}, i.e., with GPT-4 with short prompts on the 10 overall runs (5 for two accounts each). A checkmark indicates a correct answer. FN and FP indicate that the answer contains a false negative or false positive, respectively. $^*$ indicates that the answer changed after the system was asked to rethink the answer. \label{tab:results_gpt4_long_complete}}
\end{center}
\end{table}

\section{Mock privacy policy sample (spreadsheet)}\label{sec:appendixB}

\begin{landscape}
\begin{figure}
  \includegraphics[height=.65\textheight]{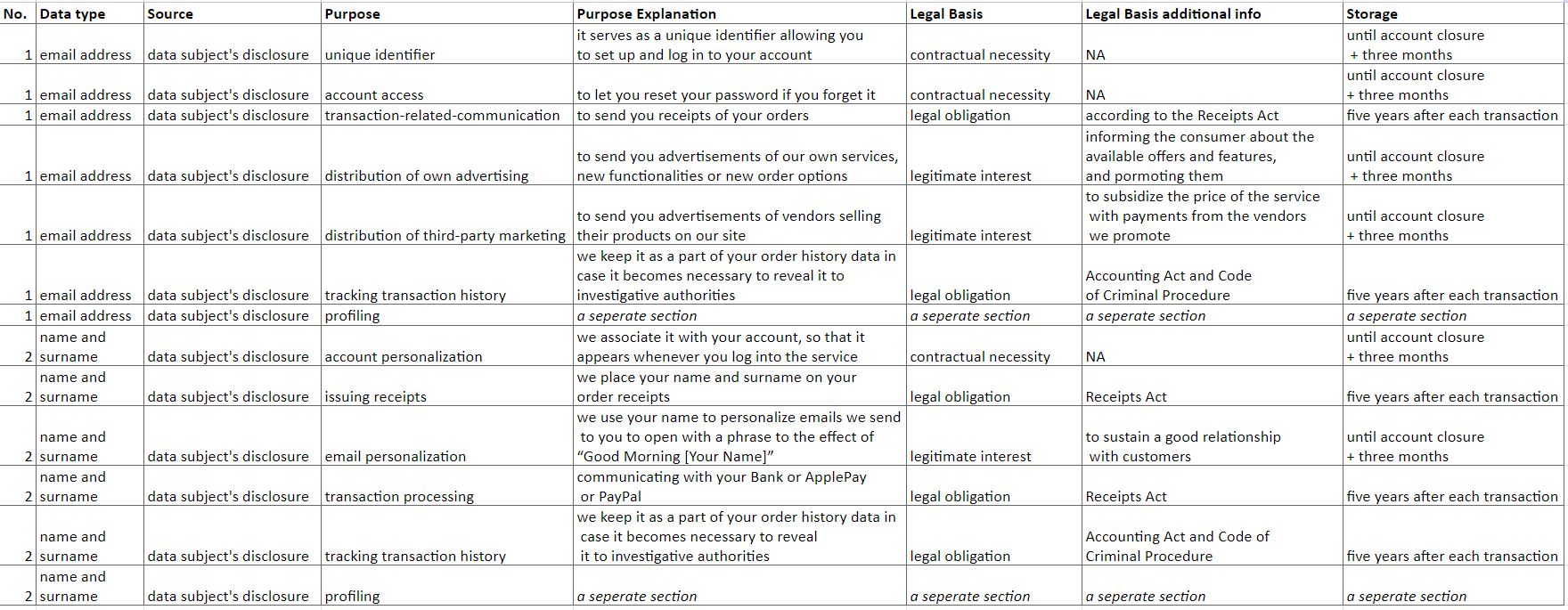}
  \caption{Mock privacy policy}
  \label{fig:pp}
\end{figure}
\end{landscape}

Figure \ref{fig:pp} presents a sample of the mock privacy policy. It includes two data types (i.e., email address, name and surname) their specified sources, purposes for processing, explanations of these purposes, legal basis and further explanations, and information about storage.  

\section{A sample of the mock privacy policy (solid text)}\label{sec:appendixC}

\input{sample_privacy}

\end{document}

%% file: sample_privacy.tex

\begin{center} 
\textbf{THIS IS A MOCK POLICY OF “ORDEROO INC.” 
A COMPANY SIMILAR TO DELIVEROO/ DOORDASH /JUST EAT/ UBER EATS etc.}
\end{center}

\begin{center} 
\textbf{ORDEROO INC. PRIVACY POLICY }
\end{center}
We process your personal data in the following way:

\underline{1. Your email address.} You provide us with your email address when registering for the service. We use your email address for the following purposes: unique identifier, it serves as a unique identifier allowing you to set up and log in to your account (contractual necessity); account access, to let you reset your password if you forget it (contractual necessity); transaction-related-communication, to send you receipts of your orders (legal obligation: to issue receipts, according to the Receipts Act); distribution of own advertising, to send you advertisements of our own services, new functionalities or new order options (legitimate interest: informing the consumers about the available offers and features, and promoting them); distribution of third-party marketing, to send you advertisements of vendors selling their products on our site (legitimate interest: to subsidize the price of the service with payments from the vendors we promote); tracking transaction history, we keep it as a part of your order history data in case it becomes necessary to reveal it to investigative authorities (legal obligation: Accounting Act and Code of Criminal Procedure); we use the domain name part of your email when profiling (see the separate section at the bottom of the document). We share your email address with Cloud711 (processor), for the purpose of data storage and backup, i.e., storing our IT operations on their servers (legitimate interest: lowering the cost of operation and keeping the data safe); Microsoft (processor) for the purpose of facilitating communication, i.e., sending our own emails (legitimate interest in outsourcing the operation of email servers and protocols); CoolAccountants (processor) for the purpose of accounting, i.e., reviewing our financial records and keeping them in order (legal obligation: Accounting Act); FraudDetectors (processor) for the purpose of fraud detection (legitimate interest: not becoming a victim of fraud). We do not share your email address with recipients choosing their own purposes of processing (controllers). We store your email for as long as you’re using our services, i.e., until you delete your account, PLUS THREE MONTHS. For the purposes required by the Receipts Act, Accounting Act, and the Code of Criminal Procedure, we store your email address for five years after your last transaction.